\documentstyle[12pt,epsf]{article}
\newlength{\abstractwidth}
\renewcommand{\thefootnote}{\fnsymbol{footnote}}
\renewcommand{\thanks}[1]{\footnote{#1}} % Use this for footnotes
\newcommand{\starttext}{
\setcounter{footnote}{0}
\renewcommand{\thefootnote}{\arabic{footnote}}}
\renewcommand{\theequation}{\thesection.\arabic{equation}}
\newcommand{\be}{\begin{equation}}
\newcommand{\bea}{\begin{eqnarray}}
\newcommand{\eea}{\end{eqnarray}}
\newcommand{\beq}{\begin{equation}}
\newcommand{\ee}{\end{equation}}
\newcommand{\eeq}{\end{equation}}
\newcommand{\N}{{\cal N}}

\renewcommand{\a}{\alpha}

\newcommand{\tp}{\tilde p}

\newcommand{\half}{{1\over 2}}

\newcommand{\tpm}[1]{\tilde p^{\mu_{#1}}}

\def\ba{\begin{eqnarray}}
\def\ea{\end{eqnarray}}
\newcommand{\PSbox}[3]{\mbox{\rule{0in}{#3}\includegraphics{#1}\hspace{#2}}}

\def\N{{\cal N}}

\def\12{{1 \over 2}}
\def\32{{3 \over 2}}
\def\72{{7 \over 2}}
\def\92{{9 \over 2}}

\begin{document}
\renewcommand{\theequation}{\thesection.\arabic{equation}}
\begin{titlepage}
\bigskip
\hskip 3.7in\vbox{\baselineskip12pt \hbox{SU-ITP 00-07} \hbox{hep-th/0002075}}

\bigskip\bigskip\bigskip\bigskip

\centerline{\Large \bf The IR/UV Connection}
\medskip
\centerline{\Large \bf in Non--Commutative Gauge Theories}

\bigskip\bigskip
\bigskip\bigskip
\centerline{\bf Alec Matusis, Leonard Susskind and Nicolaos Toumbas
\footnote[1]{\tt alecm@stanford.edu,
susskind@sewerrat.stanford.edu, ntoumbas@leland.stanford.edu} }
\bigskip
\bigskip
\centerline{\it Department of Physics} \centerline{\it Stanford
University} \centerline{\it Stanford CA 94305-4060}
\bigskip\bigskip
\begin{abstract}

\medskip
\noindent Quantum field theory on non-commutative spaces does not
enjoy the usual ultraviolet-infrared decoupling that forms the
basis for conventional renormalization. The high momentum
contributions to loop integrations can lead to unfamiliar long
distance behavior which can potentially undermine naive
expectations for the IR behavior of the theory. These ``anomalies"
involve non--analytic behavior in the noncommutativity parameter
$\theta$ making the limit $\theta \to 0$ singular.

In this paper we will analyze such  effects in the one loop
approximation to gauge theories on non-commutative space. We will
see that contrary to expectations poles in $\theta$ do occur and lead
to large discrepancies between the expected and actual infrared
behavior. We find that poles in $\theta$ are absent in supersymmetric
theories. The ``anomalies" are generally still present, but only at
the logarithmic level. A
notable exception is non-commutative super Yang Mills theory with
16 real supercharges in which anomalous effects seem to be absent
altogether.
\end{abstract}
\end{titlepage}
\starttext \baselineskip=18pt \setcounter{footnote}{0}

%%%%%%%%%%%%%%%%%%%%%%%%%%%%%%%%%%%%%%%%%%%%%%%%%%%%%%%%%%%%%%%%%%%%%%
%%%%%%
%%%%%%%%%%%%%%%%
%%%%%%%%%%%%%%%%%%%%%%%%%%%%%%%%%%%%%%%%%%%%%%%%%%%%%%%%%%%%%%%%%%%%%%
%%%%%%
%%%%%%%%%%%%%%%%
\setcounter{equation}{0}
\section{The UV/IR Connection in Non-commutative Field Theory}
Field theories formulated on non-commutative spaces are
interesting in both  their own right as well as for their
applications to string and matrix theories \cite{1}\cite{2}. These
theories are characterized by a non-commutativity parameter
$\theta$ with dimensions of length squared. Classically and in the
tree--level approximation the behavior of the theory for momenta much
less than $\theta^{-1/2}$ is the same as for the corresponding
commutative theory. However this is not necessarily the case for
the quantum theory. The non-commutativity can lead to unfamiliar
effects of the ultraviolet modes on the infrared behavior which
have no analog in conventional quantum field theory \cite{5}.

 The origin of the
strange mixing of IR and UV effects  in non-commutative field
theory can be understood in a simple way \cite{3} \cite{4}. The
field quanta in such a theory can be thought of as pairs of
opposite charges moving in a strong magnetic field . The spatial
locations of the two charges are defined by a center of mass
position $x^i_{cm}$ and a relative coordinate $\Delta^m$. The
relative coordinate is related to the spatial momentum $p$ by
\begin{equation}
\Delta^i = \theta^{ij}p_j
\end{equation}
where $\theta$ is an antisymmetric matrix with components in the
spatial directions. In this paper we will consider the case of 3
dimensional space. Without loss of generality $\theta $ can be
taken to lie in the $(1,2)$ plane
\begin{eqnarray}
\theta^{1,2}& =&-\theta^{2,1}\equiv \theta \cr \theta^{1,3} &=&
\theta^{3,2} =0.
\end{eqnarray}
The momentum in the $(1,2)$ plane will be called $P$. We will also
use the notation $\tilde{P}_i=\theta_{ij}P^j.$ Thus a particle
moving with momentum $P$ along the $X^1$ axis has a spatial
extension of size $| \theta P|$ in the $X^2$ direction. The growth
of the size of a particle with its momentum has interesting
consequences. For example, when a quantum of momentum $P$ scatters
off a target at rest, the scattering amplitude will spread in
impact parameter space over a distance $|\theta P|$.

There are also important  and somewhat bizarre consequences for
Feynman loop integrations. Roughly speaking, when a particle of
momentum $P$ circulates in a loop it can induce an effect at
distance $|\theta P|$. The high momentum end of the integrals can
give rise to power law long range forces which are absent entirely
in the classical theory. We will call such effects ``anomalies''
although we emphasize that they do not signal an inconsistency in
the theory but rather a violation of naive expectations.

In the case of planar diagrams the effect of non-commutativity is
simple. Every diagram gets multiplied by a phase factor that
depends only  on the momenta of the external lines. Thus the
Feynman integrals are exactly as in the commutative theory. In the
non-planar case the situation is more interesting. The phase
factors now involve products of external momenta $p$ and internal
momenta $l$ in the form
\begin{equation}
e^{ i p  \theta l} =e^{ i p\tilde{l}}.
\end{equation}
If the diagram in question is divergent in the commutative theory,
the effect of the oscillating phases is typically to regulate the
diagram and render it finite. But as $P \to 0$ the phases become
ineffective and the diagram diverges at $P=0$. This is the
mechanism described in detail in \cite{5}.  We will begin by
reviewing a simple example from non-commutative $\phi^4$ theory.

\begin{figure}
\centerline{\epsfysize=1.00truein \epsfbox{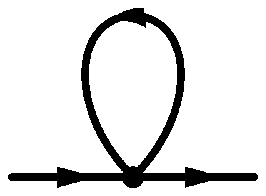}}
%\vskip -0.4 cm
 \caption{Mass renormalization correction in $\phi^4$ thoery.}
\label{diagrams}
\end{figure}

The diagram in question, fig(1), is the lowest order mass
renormalization correction to the propagator. We are interested in
the  non-planar contribution which in the commutative theory  has
the form
\begin{equation}
\int d^4l\; {1 \over l^2}.
\end{equation}
The diagram is quadratically divergent and is renormalized by a
mass counter term.

In the non-commutative case the integrand has an additional factor
$\exp (i \tilde{p}l)$ where $p$ and $l$ are the external \footnote{In
the rest of the paper, we denote by small p the 4--momentum vector
with $p_{1,2}\equiv P_{1,2}$ in the non--commutativity plane.} and loop
momenta. The integral has the form
\begin{equation}
\int d^4l\; {1 \over l^2}\,e^{ i \tilde{p}l} \sim {1 \over\theta^2 P^2}.
\end{equation}

As emphasized in \cite{5} there are some very striking features of
this result. The first is that the pole at $P=0$ arises from the
high momentum region of integration. Although we evaluated it for
the massless theory, the pole itself is independent of mass.
Furthermore this contribution to the self energy has a huge effect
on the propagation of long wavelength particles. The on-shell
condition or dispersion relation becomes
\begin{equation}
p_0^2=p_3^2 + P^2 + c{1 \over\theta^2 P^2}
\end{equation}
where $c $ is proportional to the coupling constant. Thus, as
discussed in \cite{5}, the behavior of the non-commutative theory
below the non-commutativity scale seems to be nothing like the
commutative theory. In this case the low momentum end of the
spectrum is completely removed from the low energy theory.

Commutative gauge theories are better behaved in the UV than
commutative scalar theories. The worst divergences in pure Yang
mills theory or Yang Mills theory with fermions are logarithmic.
This naively suggests that in their non-commutative versions the
worst anomalous effects will be logarithmic in $P$. As an example
consider the vacuum polarization correction to the gauge boson
propagator. The divergences have the form
\be
\Pi \sim g^2 p^2 \log \kappa \ee where $p^2$ is the squared
four-momentum of the gauge boson. Note in particular that the mass
correction vanishes since $\Pi$ vanishes at $p^2=0$. This
situation suggests that in the noncommutative theory the worst
anomalous effect in the propagator has the form
\be
\Pi \sim g^2 p^2 \log \tilde{P}^2 \ee If this were so, the dispersion
relation of a low energy gauge boson would be unaffected  by the
non-commutativity. As we will see in the next section this is
generally incorrect.
%%%%%%%%%%%%%%%%%%%%%%%%%%%%%%%%%%%%%%%%%%%%%%%%%%%%%%%%%%%%%%%%%%%%%%%%%%%%
%%%%%%%%%%%%%%%%%%%%%%%%%%%%%%%%%%%%%%%%%%%%%%%%%%%%%%%%%%%%%%%%%%%%%%%%%%%%
%%%%%%%%%%%%%%%%%%%%%
\setcounter{equation}{0}
\section{U(1) Non--Commutative Yang--Mills}

In this section we analyze $U(1)$ Yang--Mills theory on a
non-commutative space. The classical action is given by
\begin{equation}
S = - {1 \over {4}} \int d^4x F^2,
\end{equation}
with the field strength F given by
\begin{equation}
F_{\mu\nu} = \partial_{\mu}A_{\nu} -
\partial_{\nu}A_{\mu}-ig[A_{\mu} , A_{\nu}]
\end{equation}
and
\begin{equation}
[A_{\mu} , A_{\nu}] = A_{\mu}*A_{\nu} - A_{\nu}*A_{\mu}.
\end{equation}
The $*$--product between two functions $\phi_1(x)$ and $\phi_2(x)$
is defined by
\begin{equation}
\phi_1*\phi_2(x) = e^{{i \over 2} \theta^{\mu\nu} \partial^y_{\mu}
\partial^z_{\nu}} \phi_1(y) \phi_2(z)|_{y=z=x}.
\end{equation}
The theory is invariant under non--commutative gauge
transformations
\begin{equation}
\delta_{\lambda} A_{\mu} = \partial_{\mu}\lambda -
gi(A_{\mu}*\lambda - \lambda*A_{\mu}).
\end{equation}

We may add matter fields in the theory as well. The scalar and
fermionic parts of the action that involve interactions of the
matter fields with the gauge field are given by
\begin{equation}
S_{matter} = \int d^4x i\bar{\psi} * \gamma^{\mu}D_{\mu}\psi + {1
\over 2}(D_{\mu}\Phi)^2.
\end{equation}
The covariant derivative acts on the fields as follows
\begin{equation}
D_{\mu}X = \partial_{\mu}X - ig[A_{\mu} , X].
\end{equation}
The commutator is defined through the $*$--product as before. The
matter fields are covariant under the family of gauge
transformations given in Eq(2.5).

The Feynman rules for the theory have been worked out in
references \cite{6}\cite{7}. The vertices look similar to those of
a commutative non--abelian gauge theory with all matter fields in
the adjoint of the gauge group. The structure constants are
replaced by sines of external momenta as shown in detail in
the Appendix. The Feynman rules for the ghosts are also included. The
vertices vanish when $\theta$ is taken to be zero as expected. Our
computations are done in the Feynman gauge. In this gauge the
gauge field propagator is given by
\begin{equation}
-i{g_{\mu\nu} \over p^2}.
\end{equation}
We refer the reader in \cite{6}\cite{Filk} for a derivation of the
perturbative Feynman rules and gauge fixing of the theory.

\setcounter{equation}0
\section{Photon self--energy correction}

In non-commutative gauge theory the  most serious anomalous
effects, namely those which exhibit inverse powers of $\tilde{P}$,
are associated with the 2 and 3 point functions. We will begin
with the computation of the 2-point photon self energy diagrams.
We will consider the contributions from loops involving fermions,
scalars and gauge bosons (including ghosts). It should be noted
that the individual contributions are gauge invariant. Similar
calculations have been done by Hayakawa \cite{7}. The relevant
diagrams are shown in Figures 2 and 3. We will illustrate the
procedure with the fermion loop (line (1) in Fig. 3).

\begin{figure}
\centerline{\epsfysize=1.00truein \epsfbox{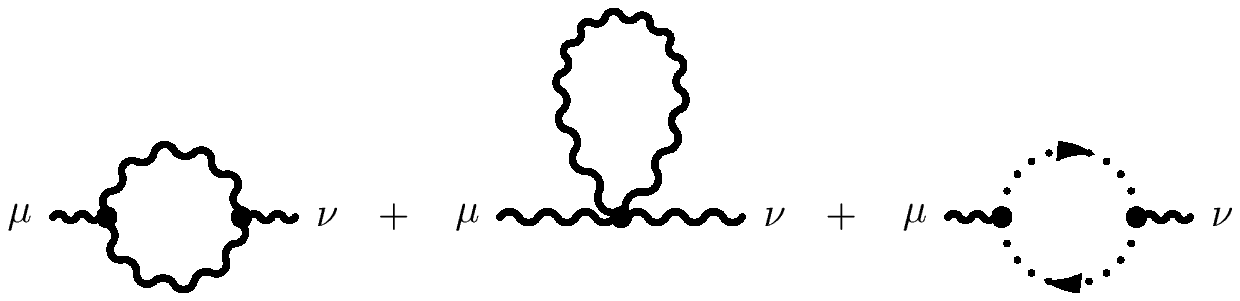}}
%\vskip -0.4 cm
 \caption{Photon self energy: gauge loop corrections.}
\end{figure}
\begin{figure}
\centerline{\epsfysize=1.50truein \epsfbox{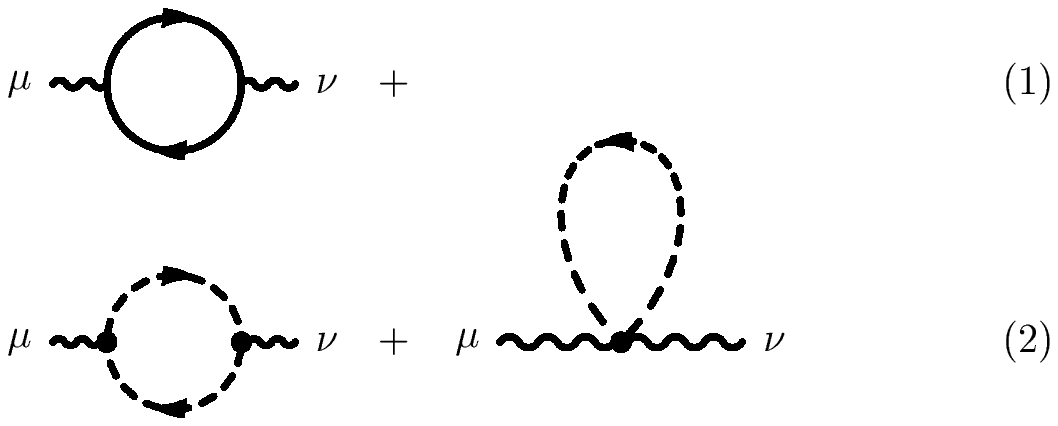}}
%\vskip -0.4 cm
 \caption{Photon self energy: matter loop corrections.}
\end{figure}

Using the Feynman rules from the Appendix we find
\be
i\Pi^{\mu\nu}_f(p)=-4g^2N_f\int {d^4l \over (2\pi)^4}{{\rm
Tr}\left[\gamma^\mu (/\!\!l - /\!\!\!p)\gamma^\nu /\!\!l\right]
\over(l-p)^2l^2}\sin^2 \left(\half \tilde p l\right), \ee where
$N_f$ is the number of Majorana fermions, each of which counts as two
fermion species.

We are interested in the contribution to the integral coming from
very high loop momentum. We therefore drop the sub-leading
dependence in the integrand and replace eq.(3.1) by
\be
i\Pi^{\mu\nu}_f(p)=-4g^2N_f\int {d^4l \over (2\pi)^4}{{\rm
Tr}\left[\gamma^\mu /\!\!l \gamma^\nu /\!\!l\right]
\over(l)^4}\sin^2 \left(\half \tilde p l\right), \ee Using
\be
\sin^2 \left(\half \tilde p l\right)=\half [1-\cos  \left( \tilde
p l\right) ], \ee we can isolate the planar and non-planar
contributions. The non-planar contribution is obtained by dropping
the first term and keeping only the $\cos$ term.
 Working out the trace,
we get \be\label{approxint} i\Pi^{\mu\nu}_f(p)= 4g^2N_f\int {d^4l
\over (2\pi)^4} {\left(2l^\mu l^\nu-g^{\mu\nu}l^2\right) \over
l^4}\; e^{i\tilde p\, l}. \ee Since
$
\int d^4l /(2\pi)^4 {1 \over l^4}\; e^{i\tilde p\,
l}=i\alpha\;(\log\Lambda-\log |\tilde p|) $ \footnote{In Euclidean
space the integral is of course real, $ \int d^4l /(2\pi)^4 {1
\over l^4}\; e^{i\tilde p\, l}=\alpha\;(\log\Lambda-\log |\tilde
p|),\;\;\alpha >0. $}, where $\Lambda$ denotes a short--momentum
cut--off and $\a$ is a positive real constant, we can rewrite
(\ref{approxint}) as
\be
i\Pi^{\mu\nu}_f(p)= 4ig^2N_f\alpha
\left(2\,\partial^\mu\partial^\nu-g^{\mu \nu}\partial^2\right)\log
|\tilde p|.
 \ee
The integral (\ref{approxint}) is finite, and the cut--off
dependence vanishes after differentiating. We note that the term
$g^{\mu\nu}/ \tilde p^2$ cancels in this expression leaving us
with \be\label{2ptpole} i\Pi^{\mu\nu}_f(p)=
-8ig^2(2N_f)\alpha\;{\tilde p^\mu\tilde p^\nu\over \tilde p^4}. \ee

 The answer is somewhat surprising. If the factor $\sin^2
\left(\half \tilde p l \right)$ were not present in eq(3.1) the
expression would be the conventional self energy diagram of the
commutative theory. Gauge invariance would be invoked to say that
any quadratic divergence is absent. Alternatively the diagram can
be Pauli Villars regulated eliminating the quadratic divergence.
The integral with the trigonometric factor is finite and well
defined. However it quadratically diverges as $\tilde{p} \to 0$.
Thus we see that there is an anomalous effect of order $\theta^{-2}$
arising out of a
diagram which in the commutative theory is quadratically divergent
by power counting but for which the divergence vanishes as a
consequence of symmetry. As noted in \cite{5} in the context of
scalar theories this type of behavior is proportional to inverse
powers of $\theta$. Evidently the limit in which $\theta \to 0$
does not smoothly tend to the commutative theory.

The physical interpretation of terms like eq(3.6) is very
interesting.
 For small non-commutative momentum, the one-loop inverse propagator is
given by
\begin{equation}
\Gamma_{\mu\nu} = i\left[(p_0^2 -p_3^2- P^2)g_{\mu \nu} - g^2c{\tilde
{p}_{\mu} \tilde {p}_{\nu} \over \tilde {p}^4}\right],
\end{equation}
where  $P$ represents the projection of the spatial momentum on
the $(1,2)$ plane. From this matrix we can read the dispersion
relation for the two physical, transversely polarized photons.
Suppose $P$ is along the 2--direction so that $\tilde{P}$ is in
the 1-direction. Then the photon polarized in the direction
perpendicular to $\tilde{P}$ satisfies the same dispersion
relation as a photon would in the commutative theory
\begin{equation}
p_0^2 =p_3^2+ P^2.
\end{equation}
However, the photon polarized along the 1-direction, parallel to
$\tilde{P}$, satisfies a different dispersion relation given by
\begin{equation}
p_0^2 =p_3^2+ P^2 + c g^2 {1 \over \theta^2 P^2}.
\end{equation}

This splitting of the polarization states of the gauge boson is
perfectly consistent with gauge invariance.  Indeed the vacuum
polarization tensor in eq.(3.6) is purely transverse which follows
from the identity $p \tilde p=0$. This effect would not be
possible without the breaking of Lorentz invariance caused by
$\theta$.

We remark at this point that since the contributions from the
scalar loops and from the gauge sector are gauge--invariant by
themselves, they individually give combinations of the form
(\ref{approxint}). The computations involving gauge bosons (Fig.
2) and scalars (line (2) on Fig. 3)
 in the loop are very similar, and
without giving explicit details, we add all three sectors together
obtaining
\be
i\Pi^{\mu\nu}(p)= 8ig^2(N_s+2-2N_f)\,\alpha\;{\tilde p^\mu\tilde
p^\nu\over \tilde p^4}. \ee This shows that in the supersymmetric
theories with an equal number of bosons and fermions, quadratic
divergences in the photon self--energy do not appear at one loop.

As we have mentioned before, the coefficient $c=8(N_s+2-2N_f)$ of
the quadratic anomalous term is proportional to the number of
bosons minus the number of fermions in the theory. In particular
it is zero in any supersymmetric model. This had to be the case
since the splitting of the two photon states would be inconsistent
with supersymmetry. In particular, the fermions in the theory
remain massless with dispersion relation
\begin{equation}
\slash \! \! \!{p} = 0.
\end{equation}
Note also that in the $\N=4$ case the SO(6) R--symmetry of the
theory insures that the scalars do not split. So it would be
impossible to achieve bose--fermi degeneracy with multiplets of
$\N=4$ supersymmetry had $c$ been non--zero.

We also note that the coefficient $c$ will vanish in softly broken
supersymmetric theories as well. In particular, the effect we find
arises from high momenta circulating the loop. Therefore, it is
independent of the mass of the loop particle. As long as the
number of fermions equals the number of bosons in the theory c
will cancel. Simple dimensional analysis shows that the only
effect of adding a mass term in the Lagrangian is to modify the
logarithmic singularities as $\tilde{p} \rightarrow 0$.

Finally, we note that $c$ becomes negative if the number of
fermions is bigger than the number of bosons. This is also the
case in the $\phi^4$ theory when we include fermions coupled to
the scalar field through Yukawa couplings. In both cases, the
theory becomes unstable at low energies.

\setcounter{equation}0
\section{Vertex Corrections}

Next we analyze 1--loop corrections to the three--photon vertex.
In Figures 4 and 5, we have drawn all 1PI Feynman diagrams that
contribute corrections to the vertex up to 1--loop order in
perturbation theory. As before, we shall see that there are
unfamiliar long distance effects arising from the region of
integration of high loop momenta.
\begin{figure}
\centerline{\epsfysize=2.80truein \epsfbox{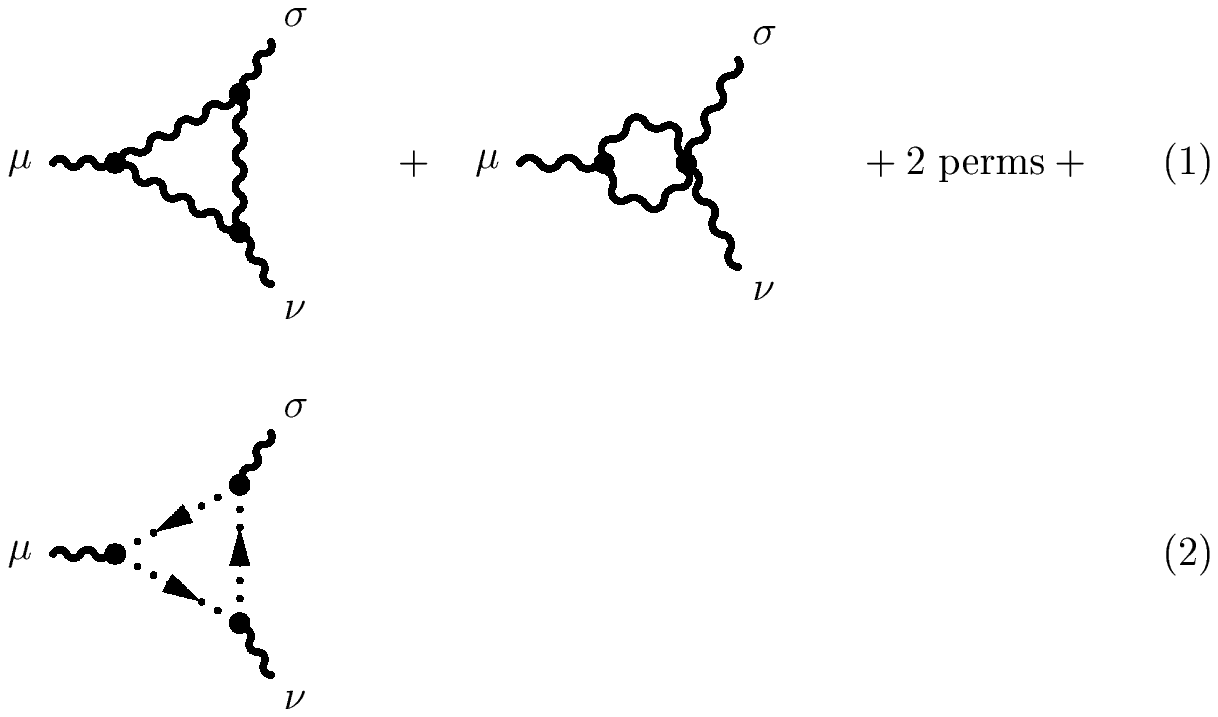}}
%\vskip -0.4 cm
 \caption{Vertex corrections: gauge sector.}
\label{diagrams}
\end{figure}
\begin{figure}
\centerline{\epsfysize=2.80truein \epsfbox{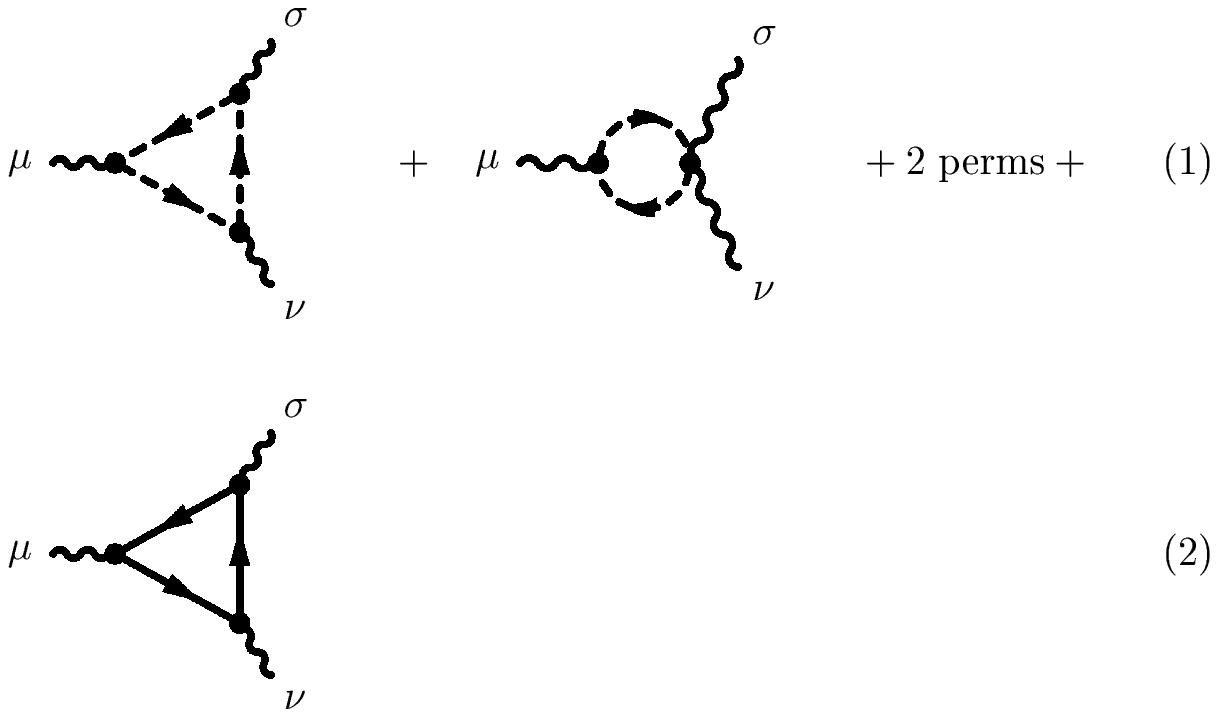}}
%\vskip -0.4 cm
 \caption{Vertex corrections: matter sector.}
\label{diagrams}
\end{figure}

Let us consider in detail the scalar graph involving cubic
vertices only (Fig. 5, 1st graph on line (1)).
 Applying the Feynman rules, we find
\begin{equation}\label{3pt}
-8ig^3\int {d^4l \over (2\pi)^4}
{(2l+p_1)^{\mu_{1}}(2l-p_2)^{\mu_{2}}(2l+p_1-p_2)^{\mu_3} \over
l^2(l-p_2)^2(l+p_1)^2}\sin{\left({\tilde{p}_1l \over
2}\right)}\sin{\left({\tilde{p}_2 l\over
2}\right)}\sin{\left({\tilde{p}_3(l+{p}_1) \over 2}\right)}.
\end{equation}
We must also impose overall momentum conservation so that
\begin{equation}
p_1 + p_2 + p_3 = 0.
\end{equation}
We have amputated the external propagators for compactness.
Eq(\ref{3pt}) has to be multiplied by the number of scalars in the
theory. Ignoring the phases for a moment, we note that the
integrand is linear in $l$ at high loop momentum
\begin{equation}
\sim \int {d^4l \over (2\pi)^4}{l^{\mu_1}l^{\mu_2}l^{\mu_3} \over
l^6}.
\end{equation}
In the commutative non--abelian theory, however, no linear
divergence arises because the integral is zero by symmetry. Thus
it is at most logarithmic in the cutoff. In the non--commutative
case the oscillating phases spoil the rotational symmetry. They
make the integral finite but they induce low momentum poles of the
form
\begin{equation}
{\tilde{p}^{\mu_1} \tilde{p}^{\mu_2}\tilde{p}^{\mu_3} \over
\tilde{p}^4}.
\end{equation}
Unlike the previous effect which was of order $\theta^{-2}$, this
is a $\theta^{-1}$ effect.

To compute the precise coefficient in front of such anomalous
terms, it is enough to consider high momentum running in the loop
ignoring external momenta in the denominators. Then the integral
becomes
\begin{equation}
-8ig^3 \cos({\tilde{p}_3 {p}_1 \over 2})\int {d^4l \over (2\pi)^4}
{8l^{\mu_{1}}l^{\mu_{2}}l^{\mu_3} \over
l^6}\sin{\left({\tilde{p}_1l \over
2}\right)}\sin{\left({\tilde{p}_2 l\over
2}\right)}\sin{\left({\tilde{p}_3l \over 2}\right)}.
\end{equation}
To carry out this integral, it is useful to express the product of
sines as a sum of exponentials. Using Eq(4.2), we can write the
product of sines as
\begin{equation}
-{1 \over 4}\left[\sin{\left({\tilde{p}_1l}\right)} +
\sin{\left({\tilde{p}_2 l}\right)} + \sin{\left({\tilde{p}_3l
}\right)}\right].
\end{equation}
We are left with a sum of three simpler integrals, and, in
addition, we can replace each sine by an exponential in the
integral. In this form, it is easy to see that the contributions
arise solely from the six non--planar graphs in the double line
notation.

To these, we must add the contributions from the scalar graphs
that involve a quartic vertex (Second graph and its two
permutations on line (1) in Fig. 5).
 The graphs produce a different
tensor structure but the phase factors are the same. There are
three such graphs from permuting the external particles among the
external lines. Each graph has to be multiplied by a symmetry
factor of $1 / 2$. Adding all four graphs together yields the
following integrals
\begin{equation}\label{vertex}
4g^3N_s \cos({\tilde{p}_3 {p}_1 \over 2})\int {d^4l \over
(2\pi)^4} {1 \over l^6}\left[{4l^{\mu_{1}}l^{\mu_{2}}l^{\mu_3} -
l^2 (l^{\mu_1}g^{\mu_2\mu_3} + {\rm perms}
})\right](e^{i\tilde{p}_1l} + e^{i\tilde{p}_2l} +
e^{i\tilde{p}_3l}) .
\end{equation}

The pure gauge sector graphs and the fermionic graphs can be
computed in the same way. One has to remember to include a
combinatorics factor of 2 for the ghosts and fermion graphs
arising from the two different cyclic orderings of the external
particles on the loop. The gauge boson quartic graph has a
symmetry factor of a $1 / 2$. Including these has the effect of
changing the coefficient in front of the integrals to
\begin{equation}
N_s + 2 - 2N_f,
\end{equation}
where $N_f$ is the number of Majorana fermions in the theory. We
see that in any supersymmetric theory the linear poles are absent.

We will now explicitly find the structure of the linear poles. To
compute the integral in (\ref{vertex}), \be\label{int1}
I^{\mu_1\mu_2\mu_3}=\int  {d^4l \over(2\pi)^4}
{\left(4l^{\mu_1}l^{\mu_2}l^{\mu_3}-l^2(l^{\mu_1}g^{\mu_2\mu_3}+
l^{\mu_2}g^{\mu_1\mu_3}+l^{\mu_3}g^{\mu_1\mu_2})\right)\over
l^6}\; e^{i\tilde p\, l}, \ee we again use
$
\int d^4l /(2\pi)^4 {1 \over l^4}\; e^{i\tilde p\,
l}=i\alpha\;(\log\Lambda-\log |\tilde p|)$. Notice that
(\ref{int1}) is convergent, and the $\Lambda$--dependent part can
thus be dropped. We can write \be\label{integral}
I^{\mu_1\mu_2\mu_3}=i\left(4\partial^{\mu_1}\partial^{\mu_2}\partial^{\mu_3}-
\Box
(g^{\mu_2\mu_3}\partial^{\mu_1}+g^{\mu_1\mu_3}\partial^{\mu_2}+g^{\mu_1\mu_2
}\partial^{\mu_3})\right)\;J(\tilde p), \ee where $J(\tilde
p)=\int d^4l /(2\pi)^4 {1 \over l^6}\; e^{i\tilde p\, l} =
-iJ_E(\tilde{p})$. The integral over Euclidean space,
$J_E(\tilde{p})$ is obtained from $J(\tilde{p})$ after performing a
Wick rotation. It satisfies
\be
\Box\,J_E(\tilde p)= {1\over \tilde p^3}\,{d\over d\tilde p}\,\tilde
p^3\,{d\over d\tilde p}\;J_E(\tilde p) = \alpha\; \log\, |\tilde p|.
\ee Solving for $J$ and substituting
\be
J(\tilde p)=-iJ_E(\tp)=-{i\alpha\over 32}\,\tilde p^2\,(4\,\log\tp-3) \ee in
(\ref{integral}) we find
\be
I^{\mu_1\mu_2\mu_3}=-2\alpha\:{\tpm{1}\tpm{2}\tpm{3}\over \tp^4}.
\ee Note that the second possible tensor structure,
$(\tpm{1}g^{\mu_2\mu_3}+\tpm{2}g^{\mu_1\mu_3}+\tpm{3}g^{\mu_1\mu_2})/P^2,$
cancelled completely, just like the photon mass term in the photon
self--energy correction (\ref{2ptpole}). This fact will turn out
to be essential in showing the gauge invariance of the S--matrix
in the next chapter.

We can now summarize the computations in this section by giving
the linearly divergent terms in the correction to the 3--point
photon vertex:
\be
\Gamma^{\mu_1\mu_2\mu_3}(p_1,p_2,p_3)=-8\alpha
g^3\,(N_s+2-2N_f)\,\cos\left({\half\tilde{p}_3
{p}_1}\right)\;\sum_{i=1}^3 {\tpm{1}_i\tpm{2}_i\tpm{3}_i\over
\tp^4_i}+\dots, \ee where the terms denoted by $\dots$ are at most
logarithmic in $\tp_1, \tp_2,\tp_3$.

The anomalous effects found in this paper and in \cite{5} are
highly nonlocal but in a particular way. The matrix elements in
eqs(3.10) and (4.14) depend depend only on the components of
momentum in the $x^1,x^2$ plane and are independent of $x^3$ and
$x^4$. Thus while nonlocal in the noncommutative directions they
are completely local in the commutative directions.

In supersymmetric theories the nonlocal $\theta^{-2}$ and
$\theta^{-1}$ effects are absent, at least at one loop. However
there are still logarithmic dependences on $P$ which are
proportional to the one-loop coefficient of the $\beta$
function\footnote{The $\beta$ function is the one controlling the
running of the t'Hooft coupling in the large $N$ limit.}
 in the corresponding commutative
non-abelian gauge theory. Once again the corresponding effects are
nonlocal only in the noncommutative directions. Theories such as
$N=4$ super Yang Mills theory with vanishing $\beta$ function seem
to be free of the non-analytic dependence on $\theta$.
%%%%%%%%%%%%%%%%%%%%%%%%%%%%%%%%%%%%%%%%%%%%%%%%%%%%%%%%%%%%%%%%%%%%%%%%%%%%
%%%%%%%%%%%%%%%%%%%%%%%%%%%%%%%%%%%%%%%%%%%%%%%%%%%%%%%%%%%%%%%%%%%%%%%%%%%%
%%%%%%%%%%%%%%%%%%%%%%%%%%%%%%%%%%%%%%%%%%%%%%%%%%%%%%%%%%%%%%%%%%%%%%%%%%%%
%%%%%%%%%%%%%%%%%%%%%%%%%
\setcounter{equation}0
\section{Gauge Invariant S--Matrix?}

In this section we show that the anomalous terms in the 2--point
and 3--point functions we computed are consistent with gauge
invariance. To check gauge invariance, we study a case involving
scattering of two gauge bosons into two fermions. We consider
1--loop diagrams only and choose the kinematic variables so that
the anomalous terms we found dominate.

All particles must be put on shell using the appropriate,
corrected dispersion relations up to 1--loop order in perturbation
theory. We choose one of the gauge bosons to be transversely
polarized. We denote its polarization by $\epsilon$ and its
momentum by p; then
\begin{equation}
\epsilon_{\mu}p^{\mu} = 0.
\end{equation}
In order to test the gauge invariance we set the polarization  of
the other gauge boson, $\theta$, equal to its momentum $q$.  The
momenta of the fermions are denoted by $(l_1 , l_2)$. Gauge
invariance requires this scattering amplitude to be zero. The tree
level diagrams contributing to the process are shown in Fig. 6.

%%%%%%%%%%%%%%

\begin{figure}
\centerline{\epsfysize=1.00truein \epsfbox{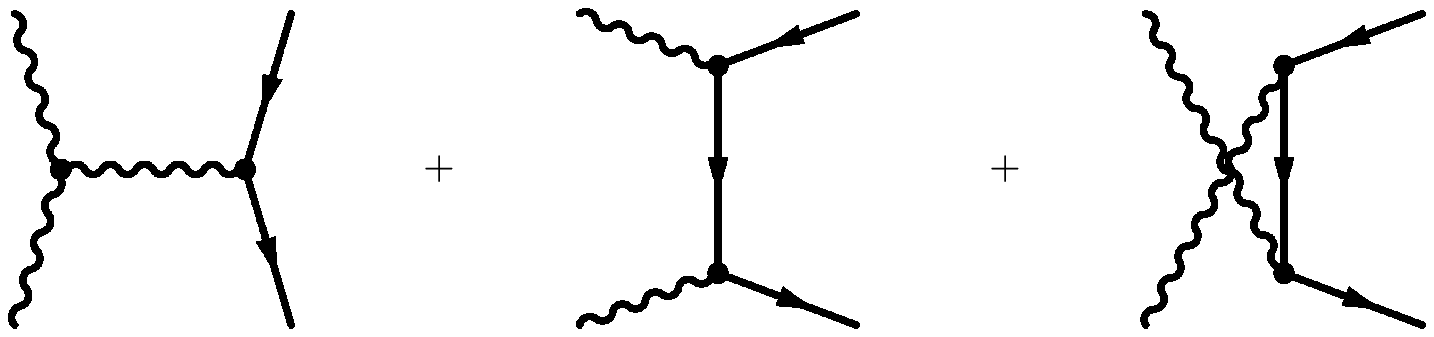}}
%\vskip -0.4 cm
 \caption{Tree level scattering diagrams.}
\label{diagrams}
\end{figure}
\begin{figure}
\centerline{\epsfysize=1.00truein \epsfbox{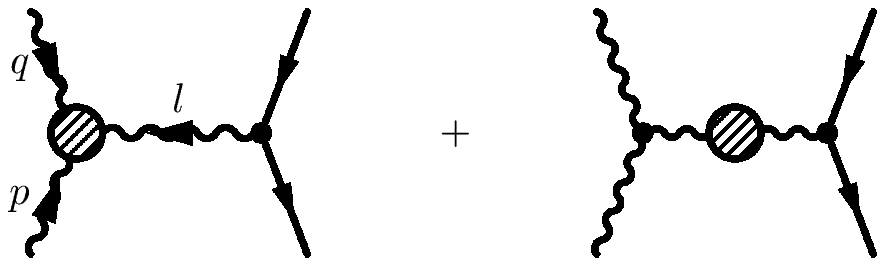}}
%\vskip -0.4 cm
 \caption{The sum of these diagrams is gauge invariant.}
\label{diagrams}
\end{figure}

%%%%%%%%%%%%%%%%%%%%%%%%%

Now we choose a specific kinematic limit. We choose $l_1 + l_2 =
-l$ so that $\tilde{l}$ is small. Furthermore, we let $l^2$ be
small so that only the s--channel diagrams are important. Two
types of one loop diagrams are important (Fig. 7), the diagrams
involving corrections to the intermediate gauge boson's propagator
and the diagrams involving corrections to the three gauge boson
vertex. Corrections to the two fermion--gauge boson vertex are at
most logarithmic in $\tilde{l}$ and, therefore, sub-leading. We
can think of the intermediate gauge boson being coupled to some
current $j_{\mu}$. The exact form of $j_{\mu}$ is not important
for our purposes.

First consider 1--loop diagrams involving the gauge boson
self--energy. In the limit when $\tilde{l}$ is small the
contribution to the amplitude becomes
\begin{equation}
i{\cal{M}}_1 = 16i\alpha g^4\sin({q\tilde{l}\over 2})
\epsilon_{\nu}q_{\mu}\left[g^{\mu\nu}(q - p)^{\rho} + g^{\nu\rho}
(p - l)^{\mu} + g^{\rho\mu} (l-q)^{\nu}\right]
{\tilde{l}_{\rho}(\tilde{l}j) \over l^4 \tilde{l}^4}.
\end{equation}
The first factor is a phase factor coming from the 3--photon
vertex. Now, using momentum conservation $-l=p+q$, $ep=0$,
$l\tilde{l}=0$ and $q^2 =0$, we see that this piece becomes
\begin{equation}
8i \alpha {(\epsilon\tilde{l})(q\tilde{l})(\tilde{l}j) \over {l^2
\tilde{l}^4}}.
\end{equation}
For small $\tilde{l}$, $\sin\left(q\tilde{l}/2\right)$ is just $q\tilde{l}/2$.
We note that the dispersion relation of a longitudinally polarized
photon remains $q^2 = 0$ since the self--energy corrections we
found are transverse.

Next we turn to diagrams involving one loop corrections to the
vertex. In the limit $\tilde{l} \rightarrow 0$, the contribution
to the amplitude becomes
\begin{equation}
i{\cal{M}}_2 = -8i \alpha
g^4{(\epsilon\tilde{l})(q\tilde{l})(\tilde{l}j) \over {l^2
\tilde{l}^4}}.
\end{equation}
This is because in this limit the vertex is dominated by anomalous
terms of the form
$\tilde{l}^{\mu}\tilde{l}^{\nu}\tilde{l}^{\rho}/\tilde{l}^4$.
Adding the two contributions together, we see that the amplitude
becomes zero as required by gauge invariance.

%%%%%%%%%%%%%%%%%%%%%%%%%%%%%%%%%%%%%%%%%%%%%%%%%%%%%%%%%%%%%%%%%%
Next we study the case when $\tilde{p} \rightarrow 0$. Again, we
study this particular case because we can isolate the possible
singular terms. The most dangerous 1--loop diagram contributing to
the scattering amplitude is the s--channel diagram involving
corrections to the 3--gauge boson vertex. All other diagrams are
sub--leading. The 1--loop contribution to the amplitude is then
given by
\begin{equation}
i{\cal{M}} = -8i \alpha
g^4{(\epsilon\tilde{p})(q\tilde{p})(\tilde{p}j) \over {l^2
\tilde{p}^4}}.
\end{equation}
We now distinguish between two cases. First we note that if
$\epsilon$ is perpendicular to $\tilde{p}$, the amplitude is zero.
Recall also from the previous section that this is the photon with
un-corrected dispersion relation. If on the other hand $\epsilon$
is along $\tilde{p}$, the amplitude is not zero but given by
\begin{equation}
i{\cal{M}}_2 = -8i \alpha g^4{(q\tilde{p})(\tilde{p}j) \over {l^2
\tilde{p}^2}}.
\end{equation}
This contribution is cancelled by the tree level graph once we use
the correct dispersion relation for the gauge boson. The tree
level graph is given by
\begin{equation}
i{\cal{M}}_3 = -2i \alpha g^2 \sin\left({q\tilde{p} \over 2}\right){(\epsilon j)(2pq) \over
l^2}.
\end{equation}
The first factor is a phase factor. We also used the following
relations: $jl=0$, $\epsilon p=0$ and $q^2=0$. Using momentum
conservation, we also find that
\begin{equation}
2pq = l^2 - p^2.
\end{equation}
Now, when the polarization of the gauge boson is along the
$\tilde{p}$ direction, the dispersion relation gets corrected as
follows
\begin{equation}
p^2 =  8 \alpha g^2{1 \over \tilde{p}^2}.
\end{equation}
Therefore, there is order a $g^4$ contribution from the
tree--level graph given by
\begin{equation}
i{\cal{M}}_3 =  8i \alpha g^4 (q\tilde{p}){(\tilde{p} j) \over l^2
\tilde{p}^2}.
\end{equation}
Adding the two together we see that to order $g^4$ the amplitude
is zero as it is required by gauge invariance.

The case involving scattering of scalars is more subtle because
the two scalar -- gauge boson vertex also contains linear
divergences in $\theta$. We defer this case for an upcoming paper
\cite{9}.

%%%%%%%%%%%%%%%%%%%%%%%%%%%%%%%%%%%%%%%%%%%%%%%%%%%%%%%%%%%%%%%%%%%%%%%%%%%%
%%%%%%%%%%%%%%%%%%%%%%%%%%%%%%%%%%%%%%%%%%%%%%%%%%%%%%%%%%%%%

%%%%%%%%%%%%%%%%%%%%%%%%%%%%%%%%%%%%%%%%%%%%%%%%%%%%%%%%%%%%%%%%%%%%%%%%%%%%
%%%%%%%%%%%%%%%%%%%%%%%%%%%%%
\section{Conclusions}

The most naive expectation about the non--commutative field
theories is that they become commutative when the
non--commutativity parameter $\theta^{ij}\rightarrow 0$ limit is
taken. In other words, star--products are replaced by ordinary
products of the fields in the Lagrangian in this limit, and one
may expect that the theory becomes commutative also at the quantum
level. As it was shown in \cite{5} this is generally not true due
to the appearance of the new divergences at low non--commutative
momenta. In the commutative theories, these divergences appear at
high momenta in the superficially divergent loop integrals, but
they can be eliminated by an appropriate choice of the
regularization scheme. In the non--commutative theories, some of
these divergences simply do not occur due to oscillating phases
associated with star--products in the vertices which make the
integrals finite. The non--commutative momenta thus play the role
of the regulator. The dependence on these non--commutative
momenta, however, does not disappear and manifests itself in the
form of the new infrared divergences at small values of
non--commutative momenta. These effects can also be characterized
as non-analytic behavior in $\theta$.

A less naive expectation would be that a non--commutative gauge
theory must be free of quadratic and linear poles at low
non--commutative momenta since the corresponding commutative
non--abelian gauge theory contains at most logarithmic
divergences. In this paper we have shown that even this
expectation does not hold, and both quadratic and linear poles
appear in a generic gauge theory. The structure of these new poles
is   consistent with gauge invariance but not Lorentz invariance.
These effects are local in the direction perpendicular to the
non--commutativity plane and completely non--local in the
non--commutative directions.

In  supersymmetric gauge theories these poles cancel between the
bosons and the fermions at the one loop level but even these
theories typically contain  logarithmic divergences at small
values of non--commutative momenta. We expect that in $\N=4$ SYM
theory even the logarithmic divergences do not occur, and thus
this theory is completely free of anomalous effects in the small
non--commutative momentum limit. This is the only theory that we
know that reduces to its commutative counterpart in the limit
$\theta^{ij}\rightarrow 0.$

\section*{Acknowledgements}
L.S. would like to thank Nathan Seiberg for numerous
discussions. We would also like to thank Steve Shenker for useful
conversations. This work was supported in part by the NSF grant 9870115.

\newpage

\section*{Appendix}
\hspace*{-1.5in}\PSbox{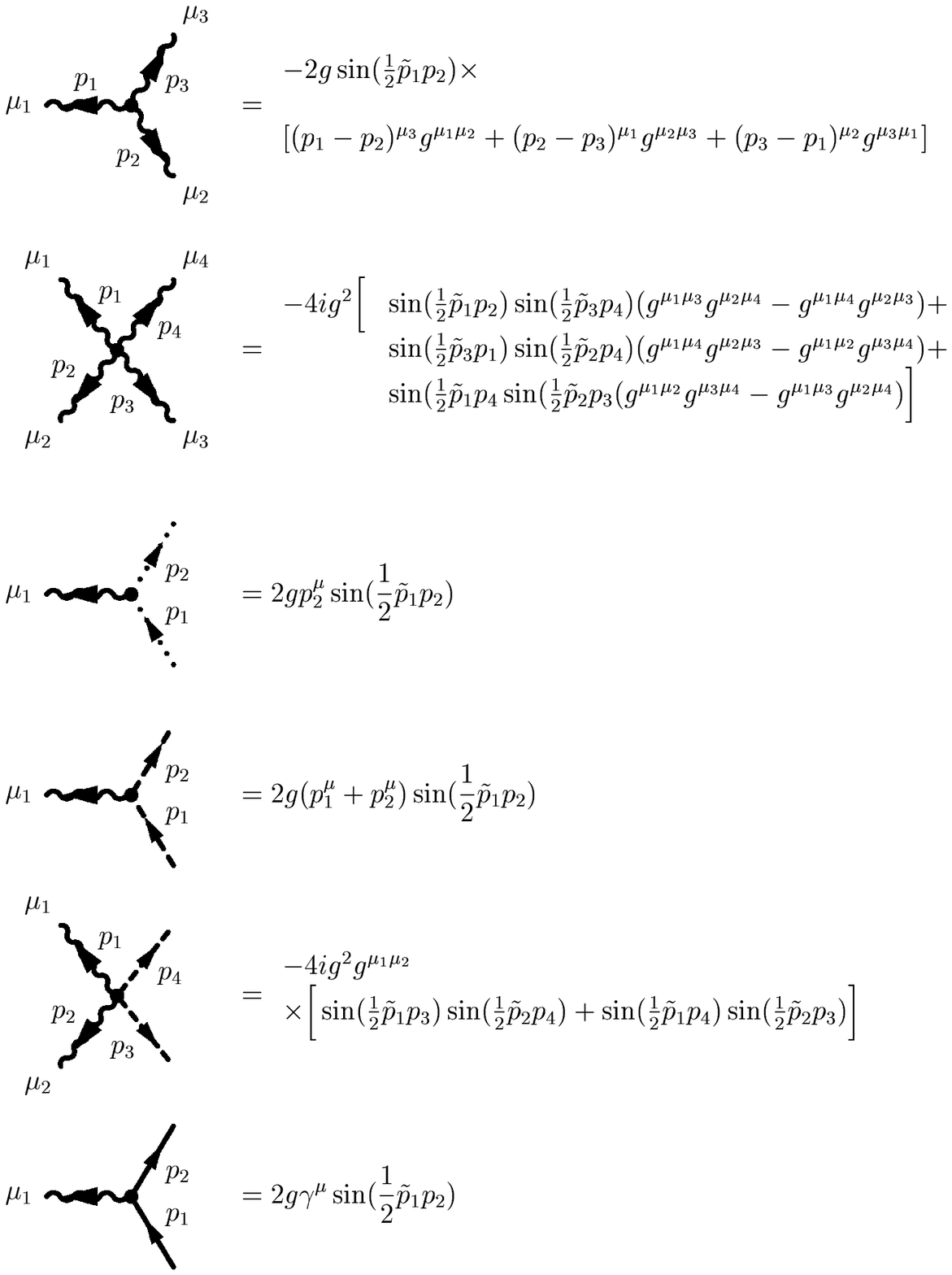 hscale=100
vscale=100}{15.1in}{9.4in}
%%%%%%%%%%%%%%%%%%%%%%%%%%%%%%%%%%%%%%%%%%%%%%%%%%%%%%%%%%%%%%%%

\end{document}